# Lu-H-N phase diagram from first-principles calculations

Fankai Xie[1,2], Tenglong Lu[1,2], Ze Yu[1,2], Yaxian Wang[1], Zongguo Wang[3], Sheng Meng[1,2,4]*, Miao Liu[1,4,5]*

[1]*Beijing National Laboratory for Condensed Matter Physics, Institute of Physics, Chinese Academy of Sciences, Beijing 100190, China*

[2]*School of Physical Sciences, University of Chinese Academy of Sciences, Beijing 100190, China*

[3]*Computer Network Information Center, Chinese Academy of Sciences, Beijing 100083, China*

[4]*Songshan Lake Materials Laboratory, Dongguan, Guangdong 523808, China*

[5]*Center of Materials Science and Optoelectronics Engineering, University of Chinese Academy of Sciences, Beijing 100049, China*

*Corresponding author: smeng@iphy.ac.cn, mliu@iphy.ac.cn*

*F. Xie and T. Lu contributed equally to this work*

## Abstract

Employing a comprehensive structure search and high-throughput first-principles calculation method on 1561 compounds, this paper reveals the phase diagram of Lu-H-N. In detail, the formation energy landscape of Lu-H-N is derived and utilized to assess the thermodynamic stability of each compound that is created via element substitution. The result indicates that there is no stable ternary structure in the Lu-H-N chemical system, however, metastable ternary structures, such as $Lu_{20}H_2N_{17}$ ($C2/m$), $Lu_2H_2N$ ($P\bar{3}m1$), are observed to have small $E_{hull}$ (< 100 meV/atom). It is also found that the energy convex hull of the Lu-H-N system shifts its shape when applying hydrostatic pressure up to 10 GPa, and the external pressure stabilizes a couple of binary phases such as $LuN_9$ and $Lu_{10}H_{21}$. Additionally, interstitial voids in $LuH_2$ are observed, which may explain the formation of

$Lu_{10}H_{21}$ and $LuH_{3-\delta}N_\epsilon$. To provide a basis for comparison, X-ray diffraction patterns and electronic structures of some compounds are also presented.

## Introduction

Lu-H-N compounds have drawn increasing attention recently as it was reported that certain Lu-H-N phase exhibits appealing observable macroscopic quantum phenomenon at ambient condition, which may revolutionarily revolt the field of condensed matter physics. The existing literature [1] has performed characterization and pointed out the compound is the $LuH_{3-\delta}N_\epsilon$ in space group of $Fm\overline{3}m$ backed by the X-ray diffraction pattern, and doped with N. The synthesis process as they stated in the article is fairly accessible as the precursors are commercially available and experimental condition, such as the pressure, temperature, etc., is not critical by current technology. However, debate soars within the community as a group of scientists are skeptical about the results, hence extra experimental confirmations[2] from multiple sources is demanded to settle down the arguments[3]. This has set off a heat wave in the study of lutetium hydrogen systems recently[4-6]. Fortunately, recent advances in high-throughput first-principles calculation along with large-scale database[7-9] makes it a feasible method to thoroughly search possible compounds within a certain chemical system, and accurately evaluate their likelihood of existence, way quicker than experimental approaches. In this paper, we tackle the stability issue of possible phases in Lu-H-N chemical systems harnessing the high-throughput calculation and a large in-house database[10], which would hopefully provide useful knowledge for the community to understand the energy landscape of the Lu-H-N compounds.

From a scientific point of view, lanthanides hydrides are worth investigating as there are several high-temperature superconductors in this phase space under high pressure. For example, Lanthanum decahydride ($LaH_{10}$) exhibits superconductivity under 252 K at the pressure of 170 GPa[11]; Yttrium-hydrogen ($YH_9$) exhibits superconductivity under 243 K at the pressure of 201 GPa[12]. For Lu-H systems, recent experiments demonstrated that $LuH_3$ undergoes a phase transiting into the superconducting states at 12.4 K under 122 GPa[13], meanwhile, it was found that the $Lu_4H_{23}$ exhibits superconductivity at 71 K under 218 GPa[14]. However, the lanthanides hydrides are less explored compared to common metals, let alone the ternary compounds within the Lu-H-N chemical space. Given that it is alleged that a certain compound within the Lu-H-N chemical space possesses superconductivity at the very much near ambient condition[1], it is necessary to have a systematic search for viable structures within the Lu-H-N system.

Leveraging the recent advance in high-throughput first-principles calculation[9, 15-18] as well as the large trove of materials data[10], in this paper, we closely investigate the possible compounds in the Lu-H-N chemical system. Such a method is fairly effective and has been demonstrated several times in searching for ternary nitrides[19], Kagome materials[20], superconductor[21], and many others, and, more importantly, the theoretical predictions have been successfully confirmed by experiments.[22, 23] Essentially, we can use the existing compounds in a database as the structure templates and perform element substitution to create new structures within the targeting chemical systems, which is Lu-H-N in this case. Then the high-throughput first-principles calculations can optimize those structures and compute the energy of each compound. By comparing the formation energies of those compounds, the formation energy landscape can be obtained, and the thermodynamic stability of each compound can be derived to justify the likelihood of the existence of those compounds quantitatively[24].

**Methodology**

In thorough explanation, those compounds which boast the least formation energy - that is, the compounds that releases the largest amount of enthalpy in their synthesis processes - are the ones that comprise the energy hull[24]. This energy hull represents the boundary line of the maximum formation energy of a chemical structure. Therefore, the hull energy designates the lowest energy state potentially attainable by any feasible combination of elements capable of forming a compound with a given chemical composition. Upon this foundation, the energy disparity between a compound and the energy hull may furnish insight into the compound's instability, or the "energy above the hull" ($E_{hull}$). $E_{hull}$ conveys the minimum amount of energy required to break down the compound into the stable compounds present on the energy convex hull.

In order to methodically assess the feasible compounds present in the Lu-H-N systems, an initial 4478 structures were constructed employing structural templates such as X-H-N, X-H-P, X-H-As, X-Li-N, X-Na-N, and X-K-N for ternary phases, and X-H, X-N, X-Li, and X-P for binary phases, resulting in 1561 newly created compounds after duplication removal. The computation was performed utilizing the Vienna Ab initio Simulation Package[25-29] along with unified input parameters for Atomly.net. Further information with regard to the detailed input parameters could be found in ref[10, 20, 21, 30]. All the structures and calculation results presented in this paper can be found on Atomly.net materials database.

After conducting high-throughput calculations, a total of 1561 structures were computed, including 712 Lu-H-N phases, 432 N-H phases, 116 Lu-H phases, and 301 Lu-N phases. The purpose of these calculations was to create a comprehensive "phase diagram" for Lu-H-N. Subsequently, we selected all the compounds with relative stability and $E_{hull}$ values less than 150 meV/atom. Those compounds were then calculated subjecting to hydrostatic pressure of up to 10 GPa with an interval of 1 GPa. Hence, the phase evolution of those compounds under pressure can be clearly captured and the phase diagrams under different pressured states are constructed and presented in this paper.

## Results and discussions

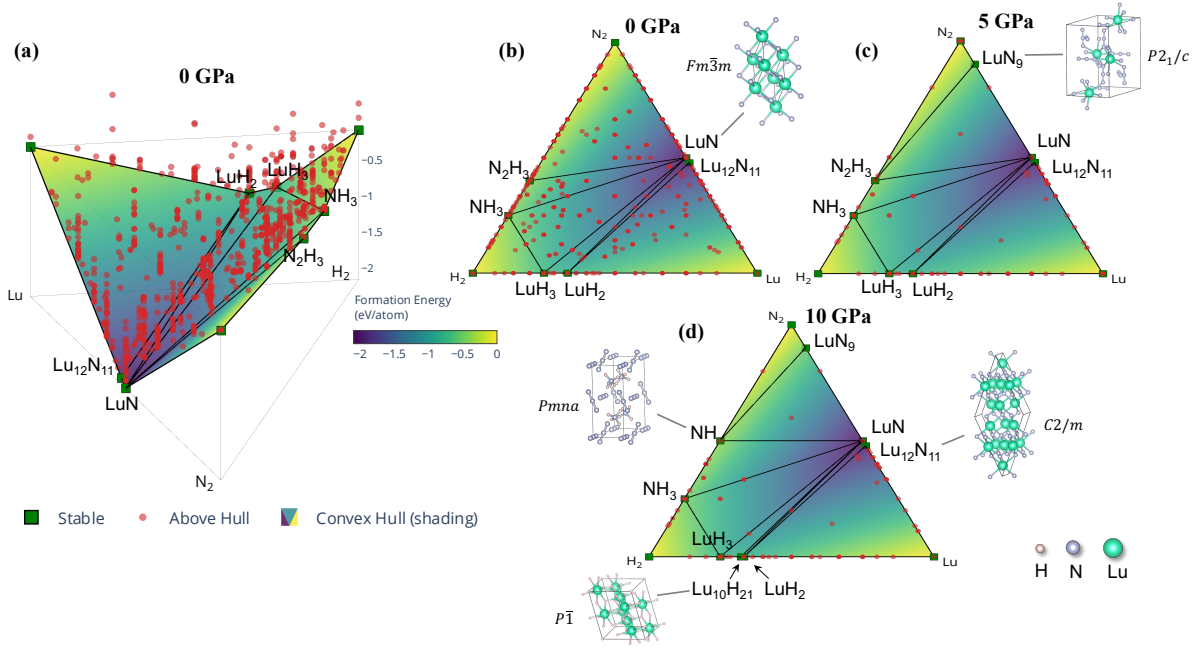

**Figure 1.** Phase diagrams of Lu-H-N chemical system in both the 3D mode and 2D mode. The phase diagrams under 0 GPa, 5 GPa, 10 GPa are calculated and presented to showcase the change of energy landscape as a function of hydrostatic pressure. On those plots, each data point shows a compound that we calculated. Stable compounds are shown in green square dots, and unstable ones are shown in red circular dots. The energy convex hull is colored based on the formation energy.

Figure 1 presents the formation energy landscape of the Lu-H-N and its evolution as a function of external pressure. The phase diagram essentially presents that, at 0GPa, there are six binary stable compounds, which are LuN ($Fm\bar{3}m$), $Lu_{12}N_{11}$ ($C2/m$), $NH_3$ ($Fm\bar{3}m$), $N_2H_3$ ($C2/c$), $LuH_2$ ($Fm\bar{3}m$), and $LuH_3$ ($P\bar{3}c1$), and there is no stable Lu-H-N ternary phase. Among all the structures we have calculated, $Lu_{20}H_2N_{17}$ ($C2/m$), $Lu_2H_2N$ ($P\bar{3}m1$), $LuH_5N_2$ ($P1$), $Lu_3H_6N$

($P2_1$), Lu$_{10}$HN$_8$ ($P\bar{1}$) are ternary metastable phases with small $E_{hull}$ (< 100 meV/atom), and the $E_{hull}$ values are listed in Table 1. When pressure is applied, we spotted some subtle changes in the phase diagram, for example, under 5 GPa, LuN$_9$ emerges as a stable phase, and under 10 GPa, Lu$_{10}$H$_{21}$ is stabilized thermodynamically. There is no thermodynamic stable Lu-H-N ternary phase according to our evaluation under external pressure up to 10 GPa.

Table 1. List of some compounds with $E_{hull}$ < 100 meV/atom at 0 GPa[10].

| Composition | Space Group | Atomly Id | $E_{hull}$ (eV/atom) | Stability |
|---|---|---|---|---|
| Lu$_{20}$H$_2$N$_{17}$ | $C2/m$ | 1000313230 | 0.036 | × |
| Lu$_2$H$_2$N | $P\bar{3}m1$ | 1000313854 | 0.082 | × |
| LuH$_5$N$_2$ | $P1$ | 1000313474 | 0.090 | × |
| Lu$_3$H$_6$N | $P2_1$ | 1000313607 | 0.095 | × |
| Lu$_{10}$HN$_8$ | $P\bar{1}$ | 1000313257 | 0.058 | × |
| NH$_3$ | $P2_13$ | 1000314459 | 0.000 | √ |
| N$_2$H$_3$ | $C2/c$ | 1000314332 | 0.000 | √ |
| LuH$_3$ | $P\bar{3}c1$ | 0000079762 | 0.000 | √ |
| LuH$_2$ | $Fm\bar{3}m$ | 1000314722 | 0.000 | √ |
| LuN | $Fm\bar{3}m$ | 3001350567 | 0.000 | √ |
| Lu$_{12}$N$_{11}$ | $C2/m$ | 1000314893 | 0.000 | √ |

Figure 2 presents the phase evolution of binary compounds with or without external pressure. Clearly, the binary phase diagrams imply that the landscape of Lu-H-N system is sensitive to pressure. For example, NH becomes stable when the pressure reaches 10 GPa. Many of the phases are not able to be found in other databases, evidencing that such a system was not thoroughly investigated previously. The existing databases such as Materials

Project[24, 31, 32], Aflow[16], and many others may miss the stable phases and thus are unable to extrapolate the correct energy convex hull for this system. Except for the stable phases, the Lu-H phase diagram shows a group of compounds that have fairly low $E_{hull}$, such as $Lu_{10}H_{21}$, $Lu_4H_7$, $Lu_3H_8$, $Lu_2H_3$, $Lu_3H_2$, $Lu_4H$, and $Lu_3H$, meaning that the Lu-H are inter-mixable and can form many likely phases at different compositions. When 10 GPa pressure is applied, a new stable phase of $Lu_{10}H_{21}$ is observed to form. However, there is no new phase that contains a higher proportion of hydrogen than the original stable phase under 10 GPa.

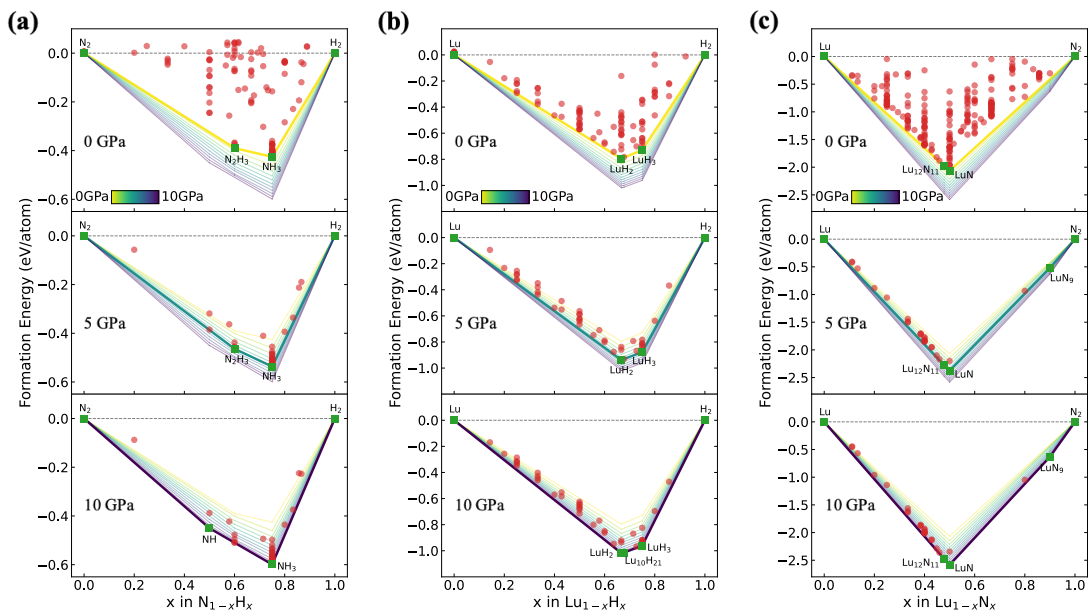

**Figure 2.** N-H, Lu-H, Lu-N binary phase diagram under 0, 5, 10 GPa separately. The compounds with $E_{hull}$ < 150 meV/atom are selected to compute formation energy under pressure > 0 GPa (a) N-H phase diagram. NH becomes a stable phase under 10 GPa, while $N_2H_3$ becomes an unstable phase. (b) Lu-H phase diagram. $Lu_{10}H_{21}$ appears to become a new stable phase under the pressure of 10 GPa. (c) Lu-N phase diagram. With the increase of pressure, Lu could combine with more N. As there appears a new phase $LuN_9$ under the pressure of 5 GPa.

The Lu-N phase diagram (Figure 2c) suggests that LuN and $Lu_{12}N_{11}$ are the thermodynamic stable phases, and there are also several low $E_{hull}$ compounds near Lu:N=1:1 composition, such as $Lu_{20}N_{17}$, $Lu_{16}N_{13}$, and $Lu_3N_2$. Under 10 GPa, it is found that as the pressure increases, the formation energy of Lu-N decreases in comparison to its value at 0 GPa, indicating that the system becomes more stable. In addition, higher pressure can also result in the appearance of a new stable phase, $LuN_9$, suggesting that high pressure could facilitate forming of nitrogen-rich Lu-N compound.

Based on our calculation, it is generally difficult to form a Lu-H-N ternary compound as the LuH and LuN do not react with each other as there is no intermediate stable phase between LuN and LuH, and the lowest $E_{hull}$ can be found are from $Lu_2HN$ and $Lu_4HN_3$, which are $E_{hull}$ =168 and 525 meV/atom, respectively, both are too unstable to warrant the existence of those compounds. The reaction calculations give that when LuH and LuN are mixed, the most likely reaction is to form $Lu_{12}N_{11}$ and $LuH_2$ to reach the more stable phases. There are also no stable or relatively stable compounds within the *Immm* space group in Lu-H-N phases based on our evaluation with or without pressure.

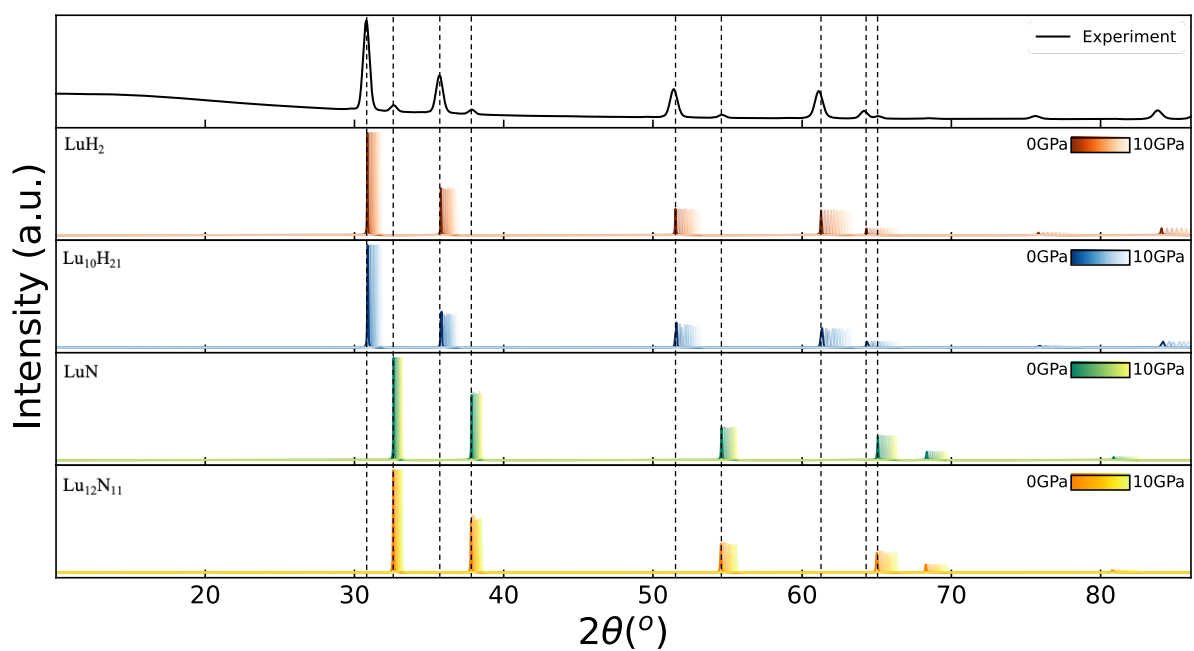

**Figure 3.** The diffraction patterns of stable structures that matches the experimental XRD[1]. Those compounds are $LuH_2$, $Lu_{10}H_{21}$, $LuN$, $Lu_{12}N_{11}$, in which $Lu_{10}H_{21}$ is basically the $LuH_2$ with addition interstitial H and $Lu_{12}N_{11}$ is LuN with N vacancy.

Figure 3 plots the compounds that share the same diffraction patterns as experimental observations among all the stable phases. Based on these patterns, it appears that $LuH_2$, $Lu_{10}H_{21}$, LuN, and $Lu_{12}N_{11}$ are likely to exist. Further investigation reveals that $Lu_{10}H_{21}$ is actually the same structure as $LuH_2$ but with H interstitials, and $Lu_{12}N_{11}$ is the same structure as LuN but with N vacancies. This suggests that increasing hydrostatic pressure may introduce H interstitials into $LuH_2$. At low doping concentrations, the H interstitials do not significantly change the volume of $LuH_2$ and may even contract the lattice parameter slightly. However, it is difficult to observe the $Lu_{10}H_{21}$ phase directly from the XRD as the volume of $Lu_{10}H_{21}$ is only 0.4% less than that of $LuH_2$. Similarly, the results suggest that N vacancies can easily form in LuN, as they are energetically favored and do not add too much strain energy. This may help explain the experimental formation of $LuH_{3-\delta}N_\epsilon$ with additional dopant, as the $LuH_2$ favors interstitial sites and can accommodate extra intercalants at low dopant concentration. When pressure is applied, the XRD peaks of $LuH_2$ shift drastically more than that in LuN, meaning the LuN is elastically stiffer than $LuH_2$. The Young's moduli of $LuH_2$ and LuN from our calculation are 148.5 GPa and 338.1 GPa, respectively.

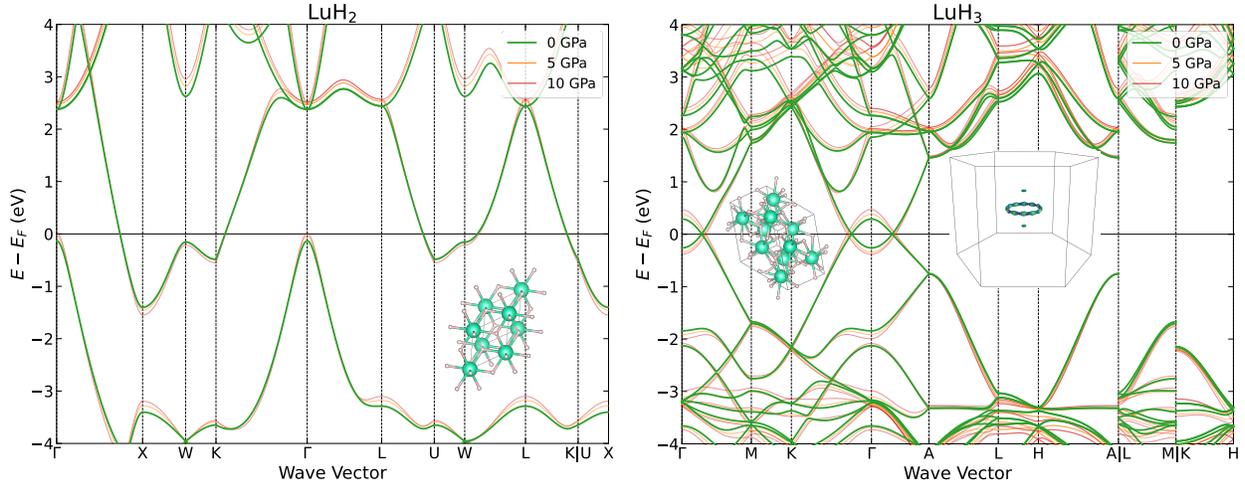

**Figure 4.** Energy bands of LuH$_2$ ($Fm\bar{3}m$) and LuH$_3$ ($P\bar{3}c1$) under equilibrium and external pressure of 5 GPa and 10 GPa from first-principles calculation at the GGA-PBE level.

Figure 4 exhibits the calculated energy bands of LuH$_2$ ($Fm\bar{3}m$) and LuH$_3$ ($P\bar{3}c1$). Based on the energy band structures, LuH$_2$ is a metallic phase and the trigonal LuH$_3$ ($P\bar{3}c1$) can be classified as a node-line semimetal when spin-orbit coupling (soc) is not considered. As shown in Figure 4, trigonal LuH$_3$ exhibits a clean electronic structure with an electron and hole pocket intersecting right at the Fermi level. Band inversion happens between |Lu$^+$, d$_{xz}$/d$_{yz}$⟩ and |H$^{1-}$, s⟩ and a nodal line is formed and protected by the glide-plane symmetry. The fact that the density of states is vanishing at the Fermi level makes LuH$_3$ an ideal system to study fundamental physics of topological node-line semimetals. With eliminated contributions of the topologically trivial bands, it is a unique platform to investigate manifestations of electron-hole interactions such as the linear frequency dependent optical conductivity[33], and the hot carrier dynamics etc[34]. Moreover, pressure-dependent band structure indicates that the nodal line is robust under hydrostatic pressure, yet it can serve as a manipulation of the Fermi velocity of the linearly dispersed bands.

## Conclusions

The recent advancement of high-throughput calculations has greatly improved the exploration and analysis of compound spaces with efficiency and accuracy. The focus of this particular study was on investigating the different phases within the chemical space of Lu-H-N, resulting in the formation energy landscape. Hydrostatic pressure was also taken into consideration to simulate experimental conditions. It is found that there is no thermodynamic stable Lu-H-N ternary phase with or without external pressure up to 10 GPa. $Lu_{20}H_2N_{17}$ ($C2/m$), $Lu_2H_2N$ ($P\bar{3}m1$), $LuH_5N_2$ ($P1$), $Lu_3H_6N$ ($P2_1$), $Lu_{10}HN_8$ ($P\bar{1}$) are ternary phases with small $E_{hull}$ (< 100 meV/atom), hence they are the potential metastable ternary phases. Additionally, the binary phase diagram shows that Lu-H compounds are inter-mixable and may have various structures due to the numerous low $E_{hull}$ Lu-H compounds calculated. The $LuH_2$ has interstitial sites, and thus can accommodate lightly doped intercalants, leading to the formation of $Lu_{10}H_{21}$ and N-doped $LuH_2$. The XRD compassion revealed that the experimental sample is likely to be composed of $LuH_2$, LuN, and dopped phases of these two. Electronic structure analysis indicated that the $LuH_2$ is a metallic phase and $LuH_3$ is a nodal line semimetal. Overall, this work aims to provide valuable insights into the synthesizability of Lu-H-N compounds, and the developed algorithm can be further utilized to explore the vast territory of inorganic compounds and to accelerate the discovery of new materials.

## Acknowledgment

We would acknowledge the financial support from Chinese Academy of Sciences (Grant No. CAS-WX2023SF-0101, and XDB33020000) and National Key R&D Program of China (2021YFA1400200, and 2021YFA0718700). The computational resource is provided by the

Platform for Data-Driven Computational Materials Discovery of the Songshan Lake laboratory.